# Spin pumping and spin torque in interfacial tailored $Co_2FeAl/\beta$-Ta layers

Ankit Kumar,[1,*] Rahul Gupta,[1] Sajid Husain,[1,2] Nilamani Behera,[1] Soumyarup Hait,[2] Sujeet Chaudhary,[2] Rimantas Brucas,[1] Peter Svedlindh,[1,†]

[1] *Department of Engineering Sciences, Uppsala University, Box 534, SE-751 21 Uppsala, Sweden*
[2] *Thin Film Laboratory, Department of Physics, Indian Institute of Technology Delhi, New Delhi 110016, India*

The Heusler ferromagnetic (FM) compound $Co_2FeAl$ interfaced with a high-spin orbit coupling non-magnetic (NM) layer is a promising candidate for energy efficient spin logic circuits. The circuit potential depends on the strength of angular momentum transfer across the FM/NM interface; hence, requiring low spin memory loss and high spin-mixing conductance. To highlight this issue, spin pumping and spin-transfer torque ferromagnetic resonance measurements have been performed on $Co_2FeAl/\beta$-Ta heterostructures tailored with Cu interfacial layers. The interface tailored structure yields an enhancement of the effective spin-mixing conductance. The interface transparency and spin memory loss corrected values of the spin-mixing conductance, spin Hall angle and spin diffusion length are found to be $3.40 \pm 0.01 \times 10^{19}\ m^{-2}$, $0.029 \pm 0.003$, and $2.3 \pm 0.5\ nm$, respectively. Furthermore, a high current modulation of the effective damping of around 2.1 % has been achieved at an applied current density of $1 \times 10^9\ \frac{A}{m^2}$, which clearly indicates the potential of using this heterostructure for energy efficient control in spin devices.

* Ankit.Kumar@angstrom.uu.se
† Peter.Svedlindh@angstrom.uu.se

# I. INTRODUCTION

Spin current manipulation of the magnetization state in magnetic heterostructures employing spin transfer torque is an effective way to achieve energy efficiency in conjunction with high endurance and non-volatility for magnetic random access memory (MRAM) and spin logic devices [1]. Gallagher *et al.* [2] reported the first MRAM operated by Oersted magnetic field switching of the magnetic state. However, Oersted field magnetic switching requires high current density, which consequently yields high Joule heating [3]. A more energy efficient way of magnetization manipulation for spin logic devices is the use of spin transfer torque (STT). Recently, L. Liu *et al.* [4] reported the spin Hall effect (SHE) induced STT switching of the ferromagnetic (FM) layer in magnetic tunnel junctions. In case of a SHE operated device, an in-plane charge current density $J_C$ in a device including a high spin orbit coupling (SOC) nonmagnetic (NM) layer generates a transverse spin current density, $J_S$. The efficiency of the SHE is governed by the spin Hall angle (SHA) of the NM layer; $\theta_H = \sigma_{xy}^s/\sigma_{xx}^c$, where $\sigma_{xx}^c$ is longitudinal charge conductivity and $\sigma_{xy}^s$ is spin Hall conductivity, which can also be expressed as $\theta_H = J_S/J_C$. The intrinsic Berry curvature contribution and the extrinsic contributions due to skew scattering and side jump govern the SHE mechanism [5]. The spin current density exerts a torque on the adjacent ferromagnetic layer in NM/FM heterostructures by the *s-d* exchange coupling [5]. The effectiveness of STT magnetization manipulation depends on the SHA of the NM layer as well as the interface transparency and spin memory loss (SML) at the FM/NM interface. Therefore, for STT spin devices the prime requirements are (i) large SHA of the NM layer, (ii) low magnetic damping of the FM layer, which reduces the critical current density requirement for manipulation of the magnetization, and (iii) high interfacial transparency and low SML; which will result in large spin angular momentum transfer across the interface. Large SHAs have been reported in 2D transition metal dichalcogenides and topological insulators; however, these are challenging to fabricate and their stability is also an issue [6]. Large SHAs have also been reported for 5d elements, e.g. Pt ($\theta_H = 0.07 - 0.3$) [7], high resistive $\beta$-Ta ($\theta_H = 0.03 - 0.35$) [8] and A15 disordered $\beta$-W ($\theta_H = 0.07 - 0.5$) [8]. The large variation in reported SHAs in studies of NM/FM systems is due to significant variations of the interfacial parameters; therefore, the mechanisms that produce the SHE is still fiercely debated. Liu *et al.* [4] have reported a giant SHE in $\beta$-Ta. Tantalum with a less than half-filled 5d shell is anticipated to exhibit a negative spin Hall angle [9]. Therefore, it would be of interest to understand the STT efficiency in terms of interfacial parameters of this system interfaced with a low damping Heusler compound. One of the most prominent Heusler compounds is Co$_2$FeAl (CFA), which exhibits low Gilbert damping ($\alpha_G$) and high Curie temperature in its well-ordered L2$_1$ phase [10]. Recently, we have reported $\alpha_G = 7.6 \times 10^{-4}$ and $\alpha_G = 1 \times 10^{-3}$ in partially L2$_1$ and B2 ordered as-deposited CFA thin films [11-12]. In a subsequent study, we also investigated $\beta$-Ta interfaced with different thicknesses of CFA with respect to the enhancement of the damping parameter employing spin pumping measurements [13]. The $\beta$-Ta/CFA bilayers exhibited low intrinsic Gilbert damping, $\alpha_G \leq 1.1 \times 10^{-3}$, high interface transparency, $T = 64\%$, and high spin pumping efficiency [13]. In the present study, $\beta$-Ta/CFA bilayers are further studied using interface tailoring; with and without using Cu as an intermediate interface layer. We observe that the interfacial Cu layer enhances the interfacial spin mixing conductance. The high value of the current induced modulation of the effective Gilbert damping ($\alpha_{eff}$) in $\beta$-Ta/CFA, makes this material combination a promising candidate for MRAM and spin oscillator devices.

## II. EXPERIMENTAL DETAILS

CFA thin films were deposited on Si/SiO$_2$ substrates using an ion beam sputtering system (Nordiko-3450). To remove surface contamination prior to deposition, the substrates were heat treated at 620 °C for 2 hours. The base pressure, working pressure, and growth temperature were kept at $7 \times 10^{-7}$ Torr, $2.4 \times 10^{-4}$ Torr and 300 °C, respectively. The grown CFA layers were interfaced with 6 nm of $\beta$-Ta and finally capped with a 2nm Cr layer. Five samples were deposited using interfacial Cu layers with different thickness ($t_{Cu}$); Si/CFA(7nm)/Cu($t_{Cu} = 0, 0.5, 1, 2, 3, 7$ nm)/β-Ta(6nm)/Cr(2nm). These samples are hereafter referred to as Cu(0), Cu(0.5), Cu(1), Cu(2), Cu(3) and Cu(7). In another batch, the grown CFA layers were interfaced with different thicknesses ($t_{Ta}$) of β-Ta layer: $t_{NM} = 2, 4, 6$ and 8 nm. Details of the growth technique and magnetic properties have been reported in our previous work [13]. In the previous study we established that the CFA/$\beta$-Ta bilayer devoid of an interfacial dead layer exhibited 68% interface spin transparency, which is even higher than for Pt interfaced devices.

The layer thicknesses, densities and surface/interface roughness were obtained using X-ray reflectivity (XRR) measurements. The scans covered the 2θ range 0°– 8°, and the XRR results were analysed using the PANaltical X'pert Reflectivity software package with a combined genetic and segmented algorithm model.

In-plane broadband FMR measurements were performed using a coplanar waveguide, where a lock-in amplifier detection technique was used. A pair of homemade Helmholtz coils generating a low-frequency (211.5 Hz) and low-amplitude magnetic field (0.25 mT) was used to modulate the microwave signal, which was detected by the lock-in amplifier. Each measurement was performed varying the dc magnetic field while keeping the microwave frequency constant. FMR spectra were recorded in the frequency range from 8 to 19 GHz in steps of 1 GHz with the dc magnetic field applied along the easy axis of the samples. Out-of-plane broadband FMR measurements were performed employing a vector network analyzer (VNA); details of this system are presented in Ref. [12]. In field sweep mode, keeping frequency constant, the magnetic field dependence of the complex transmission parameter $S_{21}$ was recorded.

The spin-torque (ST) FMR spectra were recorded on 20 μm (width, $w$) × 100 μm (length, $l$) patterned CFA/Cu($t_{Cu} = 0, 1, 7$ nm)/β-Ta structures employing a 150-μm pitch GGB Industries picoprobe. The measured resistances of CFA/Cr, CFA/Ta/Cr, CFA/Cu(3)/Ta/Cr and CFA/Cu(7)/Ta/Cr patterned structures are $R_{CFA/Cr} = 845$ Ω, $R_{CFA/\beta-Ta/Cr} = 452$ Ω, $R_{CFA/Cu(1)/\beta-Ta/Cr} = 350$ Ω and $R_{CFA/Cu(7)/\beta-Ta/Cr} = 62$ Ω, respectively. Using a parallel resistance network model the estimated resistance of the Ta layer is $R_{\beta-Ta} = 972$ Ω.

In ST-FMR measurements, the microwave current ($I_{rf}$) was injected along the sample length, as indicated by the arrow in Fig. 5(b). The ST-FMR spectra were recorded by scanning the in-plane magnetic field at 45° ($\varphi$) with respect to the direction of $I_{rf}$ at different constant frequencies ranging from 4 to 11 GHz. The ST-FMR setup uses an ultralow-noise signal generator from Rohde and Schwarz (SMF 100A) with 1–43.5 GHz frequency range, and $1 \times 10^{-6} - 0.1$-W power range with options for amplitude, phase and frequency modulation. These measurements used an internal amplitude modulation technique, where 50% internal amplitude modulation of the microwave signal at 999 Hz was used. The modulated signal was detected using a lock-in amplifier (Model SR 830 DSP). The applied power was kept at 15 dB during measurements (for more details of the setup, see Ref. [14]). We would also like to remark that the non-uniformity of the microwave power inside the patterned bar is negligible for the samples studied in this work.

## III.  RESULTS AND DISCUSSION

### A. X-ray reflectivity: Interface roughness

Figure 1 shows the XRR experimental spectra (symbols) along with the simulated spectra (lines). The simulated results, layers thickness and roughness, are presented in Table 1. The simulated results indicate sharp interfaces in all studied heterostructures.

### B. Spin pumping: ferromagnetic resonance measurements

The precession of the magnetization generates a transverse spin current to the FM/NM interface which can be expressed as $j_z^x(z) = \frac{\hbar}{4\pi} G_{eff}^{mix}(z) \left(m \times \frac{dm}{dt}\right)_x$, where $x$ and $z$ correspond to the directions of the spin polarization and the spin current [15-16], respectively. $G_{eff}^{mix}(z)$ is the position dependent effective spin mixing conductance, which can also be expressed as $G_{eff}^{mix}(z) \equiv G_{int}^{mix} \parallel (G_{SML}^I + G_{SHE}^B)$; where $G_{int}^{mix}$ is the FM/NM interfacial spin mixing conductance, $G_{SML}^I$ the spin conductance due to interfacial spin memory loss and $G_{SHE}^B$ is the spin conductance of the NM SHE layer [17].

$G_{int}^{mix}$ governs the flow of spin angular momentum across the FM/NM interface per unit area, which can be expressed, considering the Schep correction, as [18]

$$\frac{1}{G_{eff}^{mix}} = \frac{1}{G_{int}^{mix}} - \frac{1}{2G_{Sh}} + 2\frac{\lambda_{sd}/\sigma}{\tanh(t_{NM}/\lambda_{sd})}, \qquad (1)$$

where $G_{Sh} \cong \frac{e^2}{h}\left(\frac{k_F}{2\pi}\right)^2$ is the Sharvin conductance, $k_F$ is the Fermi wave vector and $G_{int}^{mix} = G_{Sh} - \sum r_{mn}^\uparrow (r_{nm}^\downarrow)^*$. Here $r_{mn}^\uparrow$ ($r_{nm}^\downarrow$) is the scattering amplitude for up (down) spins, and $t_{NM}$, $\lambda_{sd}$ and $\sigma$ are the thickness, spin diffusion length and conductivity of the NM layer, respectively. Simplifying the equation for $G_{int}^{mix}$ under the assumption that $\sum r_{mn}^\uparrow (r_{nm}^\downarrow)^* \equiv \frac{1}{(2\pi)^2}\frac{e^2}{h}\int r^\uparrow(k)(r^\downarrow(k))^* d^2k \cong 0$, the equation can be approximated as, $G_{int}^{mix} \cong G_{Sh}$. Therefore, the ratio between the interfacial spin mixing conductance of CFA interfaced with Cu and β-Ta can be expressed as $\frac{G_{int,Cu}^{mix}}{G_{int,\beta-Ta}^{mix}} \cong \left(\frac{k_{F,cu}}{k_{F,\beta-Ta}}\right)^2$, where $k_{F,cu}$ and $k_{F,\beta-Ta}$ are the Fermi wave vectors of the Cu and β-Ta layers, respectively. The values of the Fermi wave vector for Cu and β-Ta are $k_{F,cu} = 15.5\ nm^{-1}$ (taken from Ref. [19]) and $k_{F,\beta-Ta} = 11.8$

$nm^{-1}$ (taken form Refs. [20, 21]), which give $\frac{G_{int,Cu}^{mix}}{G_{int,\beta-Ta}^{mix}} \cong 1.73$. Hence, the $G_{eff}^{mix}$ value is expected to be larger for CFA/Cu/β-Ta than for CFA/β-Ta.

To understand the spin pumping phenomenon in these films, in-plane FMR spectra were recorded with the magnetic field ($H$) applied along the easy axis of magnetization. The recorded spectra were fitted using the expression,

$$\frac{dV}{dH} = S \frac{\left[\left(\frac{\Delta H}{2}\right)^2 - (H-H_r)^2\right]}{\left[\left(\frac{\Delta H}{2}\right)^2 + (H-H_r)^2\right]^2} + A \frac{2(H-H_r)\frac{\Delta H}{2}}{\left[\left(\frac{\Delta H}{2}\right)^2 + (H-H_r)^2\right]^2}, \qquad (2)$$

where $dV/dH$ is the magnetic field derivative of the microwave absorption signal, and $S$ and $A$ are the amplitudes of the symmetric and antisymmetric contributions to the measured signal, respectively. The full width half maximum linewidth $\Delta H$ and resonance field $H_r$ were used as fitting parameters. Figure 2(a) shows typical in-plane FMR spectra (symbols) for Cu(0), and solid lines correspond to fits using Eq. (2). The observed line width ($\mu_0 \Delta H$) vs. frequency ($f$) is plotted in Fig. 2(b) for Cu(0), Cu(3) and Cu(7), and the inset in figure shows $f$ vs. $\mu_0 H_r$ for Cu(0).

The $\mu_0 \Delta H$ vs. $f$ data were fitted using the expression:

$$\mu_0 \Delta H = \frac{4\pi \alpha_{eff}}{\gamma} f + \mu_0 \Delta H_0, \qquad (3)$$

where $\alpha_{eff}$ is the effective damping parameter which for the in-plane configuration in addition to the intrinsic Gilbert damping contains contributions both from two magnon scattering (TMS) and spin pumping into the β-Ta layer. $\Delta H_0$ is the frequency independent contribution to the linewidth due sample inhomogeneity and mosaicity. Here $\gamma = 180.4$ GHz/T, obtained from analysis of the out-of-plane FMR data, was kept fixed for all fittings [13]. The determined values of $\alpha_{eff}$ are 10.65± 0.02 × 10⁻³, 10.75 ± 0.03× 10⁻³ and 11.07 ± 0.03 × 10⁻³, and the values of $\Delta H_0$ are 0.1, 0.1 and 1.5 mT for the Cu(0), Cu(7) and Cu(3) samples, respectively. The frequency vs. resonance field data was fitted using Kittel's equation corresponding to in-plane easy-axis FMR measurements, and the obtained value of the effective magnetization $\mu_0 M_{eff}$ is 1.11T, which is also consistent with our previous results [13]. The in-plane determined $\alpha_{eff}$ values are nearly equal for the Cu(0) and Cu(7) samples, while the interfacial dusting layer in Cu(3) exhibits marginally higher value of $\alpha_{eff}$. The inhomogeneous linewidth $\Delta H_0$ is approximately zero for the Cu(0) and Cu(7) films, while it is substantially larger in Cu(3). In in-plane FMR spectra, TMS contributions cannot be ignored; therefore, to conclude the effect of Cu interfacial layer TMS free line-shape parameters estimation is required.

To evade the effect of TMS, out-of-plane FMR spectra were recorded in the 10 to 24 GHz frequency range for the Cu(0), Cu(0.5), Cu(1), Cu(2), Cu(3), Cu(7) and CFA(7.5)/ β-Ta($t_{NM}$: 2, 4, 8 nm) samples. The recorded magnetic field dependence of the complex transmission parameter $S_{21}$ was fitted to the following equation:

$$S_{21}(H,t) = S_{21}^0 + Dt + \frac{1}{\tilde{\chi}_0} \frac{M_{eff}(H-M_{eff})}{(H-M_{eff})^2 - H_{eff}^2 - i\Delta H(H-M_{eff})}, \quad (4)$$

where $S_{21}^0$ is the non-magnetic contribution to $S_{21}$, $\tilde{\chi}_0$ is an imaginary function of the frequency and film thickness. The term $Dt$ accounts for a linear drift of the recorded $S_{21}$ signal and $H_{eff} = 2\pi f/\gamma\mu_0$. $M_{eff}$ and $\gamma$ can be extracted by fitting the $H_r$ vs. $f$ results to the expression $\mu_0 H_r = \frac{2\pi f}{\gamma} + \mu_0 M_{eff}$. Figure 3(a) shows the magnetic field dependence of the real and imaginary components of $S_{21}$ (symbols) together with fits according to Eq. (4) (lines) for the Cu(0) sample at 10 GHz; all spectra were fitted to obtain line-shape parameters using Eq. (4). The inset in Fig. 3 (b) shows the fit of $H_r$ vs. $f$ for the Cu(0) sample using the above mentioned expression. The determined values of $M_{eff}$ (1.1 T) and $\gamma$ (180.4 GHz/T) for all samples are matching well with our previous results [13].

The extracted $\mu_0\Delta H$ vs. $f$ data were fitted using Eq. (3) to determine $\alpha_{eff}$ (cf. Fig. 3(b)). The determined values of $\alpha_{eff}$ are $5.6 \pm 0.2 \times 10^{-3}$, $3.6 \pm 0.1 \times 10^{-3}$, $3.9 \pm 0.1 \times 10^{-3}$, $4.2 \pm 0.1 \times 10^{-3}$, $4.5 \pm 0.1 \times 10^{-3}$ and $8.0 \pm 0.2 \times 10^{-3}$ for the Cu(0), Cu(0.5), Cu(1), Cu(2), Cu(3), and Cu(7) samples, respectively. The determined inhomogeneous linewidth values are 1.6, 2.2, 1.8, 1.1, 0.4 and 0 mT for the Cu(0), Cu(0.5), Cu(1), Cu(2), Cu(3) and Cu(7) films, respectively. Similarly, the $\alpha_{eff}$ values have been determined for the CFA(7.5)/β-Ta($t_{Ta}$: 2, 4, 8 nm) films and the results have been used to calculate the SML at the CFA/β-Ta interface, as will be discussed in the forthcoming section.

The out-of-plane measured $\alpha_{eff}$ values of the Cu(0.5–3) and Cu(7) samples can provide the true characteristics of the interfacial spin pumping mechanism. The $\alpha_{eff}$ values for the Cu(0.5–3) samples are in a range of $3.6 - 4.5 \pm 0.1 \times 10^{-3}$, which clearly indicates a reduction of spin-pumping compared to the spin pumping in the Cu(0) sample due to a reduction of $G_{eff}^{mix}$. The reduced spin pumping in the Cu(0.5–3) samples compared to Cu(7) is due to the fact that the Cu interfacial layer is non-continuous in the Cu(0.5–3) samples as is also is evidenced by the high sheet resistance in these samples. Therefore, a comparison between the Cu(0) and Cu(7) samples have been made to understand the true effect of the Cu layer at the interface.

The ratio of the out-plane recorded $\alpha_{eff}$ values for Cu(7) and Cu(0) is $\alpha_{eff}^{Cu(7)}/\alpha_{eff}^{Cu(0)} \approx 1.42$, which is approximately equal to the estimated $G_{int,Cu}^{mix}/G_{int,\beta-Ta}^{mix}$ ($\approx 1.73$) values. These results suggest that the Cu interfacial layer in CFA/β-Ta is helping to enhance the interfacial spin-mixing conductance and the net effective spin pumping in the structure.

## C. Spin memory loss

Fert *et al*. [22] originally coined the spin memory loss (SML) concept at magnetic heterostructure interfaces in current perpendicular to plane giant magnetoresistance devices. Later, Rojas-Sanchez *et al*. [23] and Liu *et al*. [21] established by experimental studies and first principle calculations that the interfacial SML contributes significantly to the damping enhancement. In their combined model of spin diffusion and spin pumping, SML is defined in terms of the ratio of the interface layer thickness and interface spin diffusion length, and an estimate of the interfacial resistance is required to calculate the SML parameter, which is however difficult to estimate accurately. Further, Chen et al. [16] developed a theoretical model in terms of retarded and advanced Green functions, employing interfacial SOC to calculate the SML, where the interfacial resistance value is not required to estimate SML. According to the Chen et al. model the effective mixing conductance on the FM side including spin back flow and SML can be expressed as

$$G_{eff}^{mix} = G_{int}^{mix}[1 - (1-\delta)^2 \varepsilon], \qquad (5)$$

while the equation on the NM side is

$$G_{eff}^{mix}(NM) = G_{int}^{mix}[(1-\varepsilon)(1-\delta)]. \qquad (6)$$

In these equations, $\delta$ is the SML parameter; $\delta = 0$ for no loss and $\delta = 1$ for complete loss of spin current at the interface, and $\varepsilon = G_{int}^{mix}/\left(G_{int}^{mix} + \frac{2}{3}k_F^2 \frac{\lambda_{mf}}{\lambda_{sd}} \tanh \frac{t_{NM}}{\lambda_{sd}}\right)$ is the spin back flow factor. Here $\lambda_{mf}(= 3.7$ nm$)$ is the electron mean free path (taken from Ref. [20]). The effective Gilbert damping in the FM/NM layer is given as

$$\alpha_{eff} = \alpha_{int} + G_{eff}^{mix} \frac{g\mu_B}{4\pi M_s t_{FM}}. \qquad (7)$$

To estimate $G_{int}^{mix}$, $\delta$ and $\lambda_{sd}$ for the CFA/β-Ta interface, we have used the $t_{NM}$ dependent effective damping $\alpha_{eff}(t_{NM})$ of CFA(7.5 nm)/β-Ta($t_{NM}$) in conjunction with the CFA layer thickness ($t_{FM}$) dependent effective damping $\alpha_{eff}(t_{FM})$ of CFA($t_{FM}$)/β-Ta films, as shown in Figs. 4(a) and (b), respectively. The $t_{FM}$ dependent effective damping $\alpha_{eff}(t_{FM})$ values of CFA($t_{FM}$)/ β-Ta films have been taken from our previous work [13]. The $\alpha_{eff}(t_{NM})$ vs $t_{Ta}$ and $\alpha_{eff}(t_{FM})$ vs $1/t_{CFA}$ data has been fitted simultaneously using Eqs. (5-7). The self-consistent fitting determined parameters values are $G_{int}^{mix} = 3.40 \pm 0.01 \times 10^{19}$ m$^{-2}$, $\delta = 0.24 \pm 0.05$, and $\lambda_{sd} = 2.3 \pm 0.5$ nm for the CFA/β-Ta structures. The loss of spin current at the interface is 24 %, which is high compared to the value of 2 % interfacial spin loss reported in CoFeB/β-Ta using ISHE measurements only at 9 GHz [9]. The spin diffusion length is in good agreement with values reported by Allen et al. [20] ($\lambda_{sd} = 2.5$ nm) and Sagasta et al. [24] ($\lambda_{sd} = 2.0$ nm).

**D. Spin transfer torque ferromagnetic resonance**

In ST-FMR measurements, $I_{rf}(t)$ generates a transverse spin current by the SHE, which excites magnetization precession in the CFA layer. The precession of the magnetization results in a time dependent variation of the resistance of the CFA layer owing to the anisotropic magnetoresistance (AMR). The time dependent variation of the resistance can be expressed as $R(t) = R_0 - \Delta R_{AMR} \sin^2 \psi(t)$, where $\psi(t)$ is the angle of the magnetization $\vec{M}$ with respect to $I_{rf}$ (see Fig. 5), and $\Delta R_{AMR}$ is the change in resistance when changing the magnetization direction from parallel to perpendicular with respect to $I_{rf}$. Mixing of $I_{rf}(t)$ and $R(t)$ results in the dc voltage $V_{mix}$, known as the spin diode effect. The spin orbit torques (SOTs) can be quantitatively determined by measuring the lineshape parameters of $V_{mix}$, which can be expressed as [25]

$$V_{mix} = V_S \frac{\left(\frac{\Delta H}{2}\right)^2}{\left(\frac{\Delta H}{2}\right)^2 + (H-H_r)^2} + V_A \frac{\frac{\Delta H}{2}(H-H_r)}{\left(\frac{\Delta H}{2}\right)^2 + (H-H_r)^2}, \tag{8}$$

where $V_S$ and $V_A$ are the symmetric and anti-symmetric components, respectively, of the $V_{mix}$. $V_S$ is proportional to the out-of-plane damping-like effective torque and $V_A$ is proportional to the in-plane effective torque due to the Oersted field and the SOT field-like torque; for more details see Refs. [14, 26]. Using these symmetric and anti-symmetric components, under assumption of a perfectly transparent interface, the SHA ($\theta_{SH}$) can be calculated from the expression [25, 26]

$$\theta_{SH} = \frac{V_S}{V_A} \frac{e\mu_0 M_S t_{NM} t_{FM}}{\hbar} \sqrt{\left(1 + \frac{M_{eff}}{H_r}\right)}, \tag{9}$$

where $M_S$ is the saturation magnetization of the CFA layer.

The recorded ST-FMR spectra of the Cu(0) and Cu(7) samples were fitted using Eq. 8; some recorded spectra of CFA(7nm)/Ta(6nm) for $\pm \mu_0 H$ field sweeps together with fits are presented in Fig. 5(c). The lineshape parameters $V_S$, $V_A$, $H_r$ and $\Delta H$ were obtained by fitting the ST-FMR spectra using Eq. (8). The SHA values for CFA(7nm)/Cu($t_{Cu}$)/Ta(6nm) at different constant frequencies were calculated using Eq. 9; the results are presented in Fig. 6. The SHA values determined using Eq. 9 are different for the different samples, which clearly indicates the presence of extrinsic contributions in the calculated SHA values. Since in our samples we have negligible Rashba contributions, the only possible contribution that can enhance the $\frac{V_S}{V_A}$ value, and hence the calculated SHA, is the spin pumping contribution. As effective damping parameter is 1.42 times larger in the Cu(7) sample than in the Cu(0) sample, more spin pumping is expected in the Cu(7) sample.

To calculate accurate SHA values, the Eq. (9) estimated values must be corrected for the spin pumping contribution since the ST-FMR determined $\theta_{SH}$ only accounts for SOT contributions in the lineshape. The symmetric part of the ST-FMR spectrum can contain contributions both from spin pumping, $V_{ISHE}$, and SOT, $V_{ST-FMR}^{sym}$; $V_{Total}^{sym} = V_{ISHE} + V_{ST-FMR}^{sym}$. Therefore, the SOT weight factor in ST-FMR spectrum can be expressed as $\eta = 1/\left(1 + \frac{V_{ISHE}}{V_{ST-FMR}^{sym}}\right)$. The frequency dependent $\eta$ values were estimated using the method presented in Ref. [14], and subsequently the spin pumping corrected SHA values at different frequencies are presented in Fig. 6. The spin pumping corrected calculated SHA values of the Cu(0) and Cu(7) samples are nearly equal to each other. The obtained value of the SHA lies within a range of the reported values, as the β-Ta SHA value is expected to be smaller than for the mixed β-Ta/α-Ta phases [24, 27].

To check spin-device significance of the CFA/β-Ta structure a study of the applied dc current dependent modulation of the lineshape parameters was performed. Superimposing a dc current in the ST-FMR measurement modulates the lineshape parameters. Figures 7 shows

$\mu_0 \Delta H$ vs. $f$ and $f$ vs. $\mu_0 H_r$ at different applied $I_{dc}$. The $f$ vs. $\mu_0 H_r$ data were fitted using the in-plane Kittel equation, yielding the same $\mu_0 M_{eff}$ values ($\approx 1.0\ T$) for the Cu(0) and Cu(7) samples. The variation of $f$ vs. $\mu_0 H_r$ for different $I_{dc}$ is insignificant for both samples, indicating negligible field-like torque. The variation of $\mu_0 \Delta H$ vs. $f$ data for different $I_{dc}$ is negligible for the Cu(7) sample, while it is substantial for the Cu(0) samples.

The Cu(0) sample $\mu_0 \Delta H$ vs. $f$ data for different $I_{dc}$ were fitted using Eq. (3) to extract the $I_{dc}$ dependent effective damping $\alpha_{eff}(I_{dc})$; the results are presented in Fig. 8. Here it is important to mention that the $I_{dc}$ dependent change of $\alpha_{eff}(I_{dc})$ is negligible for the Cu(7) sample; changing $I_{dc}$ from +3.0 to −3.0 mA, the $\alpha_{eff}(I_{dc})$ values are almost constant at $5.2 \pm 0.22 \times 10^{-3}$. The insignificant change in $\alpha_{eff}(I_{dc})$ is due to the fact that only about 1% of $I_{dc}$ passes through the β-Ta layer in Cu (7). On the other hand, 46 % of $I_{dc}$ passes through the β-Ta layer in the Cu(0) sample, which is responsible for the in comparison large modulation of the effective damping in this sample as shown in Fig. 8. This value of the dc current passing through the β-Ta layer is comparable to the value of 34.5 % of $I_{dc}$ reported by Nan *et al.* [28] for Ta/Py bilayers, while being significantly larger than the value 6.8 % of $I_{dc}$ reported by Huang *et al.* [29] for Ta/CoFeB/Pt trilayers structures.

The dc current dependent modulation of the effective Giblet damping $\alpha_{eff}(I_{dc})$ is given as [30]

$$\alpha_{eff}(I_{dc}) - \alpha_{eff}(I_{dc} = 0) = \left(\frac{\sin\varphi}{(H_r+0.5M_{eff})\mu_0 M_S t_{CFA}} \frac{\hbar}{2e}\right) \frac{I_{dc}\theta_{SH}}{A_{\beta-Ta}} \frac{R_{CFA/\beta-Ta}}{R_{\beta-Ta}}, \quad (10)$$

where $A_{\beta-Ta}$ ($= 14 \times 10^{-14}\ m^2$) is the cross sectional area of the β-Ta layer and $\frac{I_{dc}\theta_{SH}}{A_{\beta-Ta}} \frac{R_{CFA/\beta-Ta}}{R_{\beta-Ta}}$ is the spin current density in the β-Ta layer.

The percentage $I_{dc}$ modulation of $\alpha_{eff}(I_{dc})$ is defined as $\frac{\alpha_{eff}(I_{dc}=0)-\alpha_{eff}(I_{dc})}{\alpha_{eff}(I_{dc}=0)} \times 100\%$. The percentage modulation of $\alpha_{eff}(I_{dc})$ at $I_{dc} = \pm 3$ mA ($J_C = \pm 9.83 \times 10^9\ \frac{A}{m^2}$) is 21% for the Cu(0) sample. The observed modulation of the effective damping in this technologically important structure is significantly larger than the values of 0.17 % and 0.13 % at $J_C = \pm 1.0 \times 10^9\ \frac{A}{m^2}$ reported by Pai *et al.* [31] in W(5nm)/CoFeB(6nm) structures and by Kasai *et al.* [32] in Pt(3.5nm)/Py(1.4nm) structures, respectively. However, the observed modulation is about half of the values of 4.8 % and 4.4% at $J_C = \pm 1.0 \times 10^9\ \frac{A}{m^2}$ reported by Tiwari *et al.* [33] in TiN/epi-Py(10nm)/β-Ta(5nm) trilayers structures and by Behera *et al.* [34] in TiN/epi-CoFe(10nm)/β-Ta(6nm) trilayers structures, respectively.

## IV. CONCLUSION

In conclusion, we have studied the spin pumping and the SHE in interface tailored CFA/β-Ta heterostructures. The Cu dusting layer at the CFA/β-Ta interface reduces the spin mixing conductance, while the interface with a continuous Cu layer enhances the spin-mixing conductance and the effective spin pumping. This result is in coherence with the understanding of the spin-mixing conductance, where an increase in the number of interfacial conductance channels in the NM layer results in an enhancement of the spin-mixing conductance. The self-consistent determined SML value is 24% at the CFA/β-Ta interface. The dc current modulation of the effective magnetic damping modulation is 2.1 % at a dc current density of $1 \times 10^9 \frac{A}{m^2}$, which clearly indicates the technological significance of the CFA/β-Ta heterostructure for energy efficient STT-based magnetic random access memories and spin oscillator devices.

**ACKNOWLEDGEMENT**

This work is supported by the Swedish Research Council (VR), grant no 2017-03799, and Olle Engkvist Byggmästare, project number 182-0365.

Table 1

| Sample | CoFe$_2$Al | | Cu | | Ta | | Cr | | Cr$_2$O$_3$ | |
|---|---|---|---|---|---|---|---|---|---|---|
| | $t$(nm) ±0.02 | Σ(nm) ±0.03 | $t$(nm) ±0.02 | Σ(nm) ±0.03 | $t$(nm) ±0.02 | Σ(nm) ±0.03 | $t$(nm) ±0.02 | Σ(nm) ±0.03 | $t$(nm) ±0.02 | Σ(nm) ±0.03 |
| Cu(0) | 7.25 | 0.23 | 0 | 0 | 7.16 | 0.23 | 1.78 | 0.10 | 0.98 | 0.10 |
| Cu(0.5) | 8.33 | 0.96 | 0.49 | 0.72 | 7.68 | 0.31 | 3.26 | 1.65 | 0.71 | 0.35 |
| Cu(1) | 6.75 | 0.59 | 1.41 | 0.97 | 7.58 | 0.19 | 1.30 | 0.16 | 2.53 | 1.10 |
| Cu(2) | 8.15 | 0.91 | 2.30 | 0.78 | 7.84 | 0.71 | 3.02 | 1.14 | 2.21 | 0.76 |
| Cu(3) | 7.25 | 0.91 | 3.53 | 0.49 | 6.39 | 0.52 | 2.69 | 0.31 | 1.93 | 0.41 |
| Cu(7) | 7.34 | 0.87 | 6.89 | 0.50 | 6.16 | 0.91 | 2.55 | 0.35 | 1.06 | 0.34 |

Table 1: XRR fitting parameters; thickness and roughness/interface width of each individual layer in the Cu(0), Cu(0.5), Cu(1), Cu(2), Cu(3) and Cu(7) samples. Here $t$ and Σ, respectively, refer to the thickness and roughness of the layers.

Figure 1.

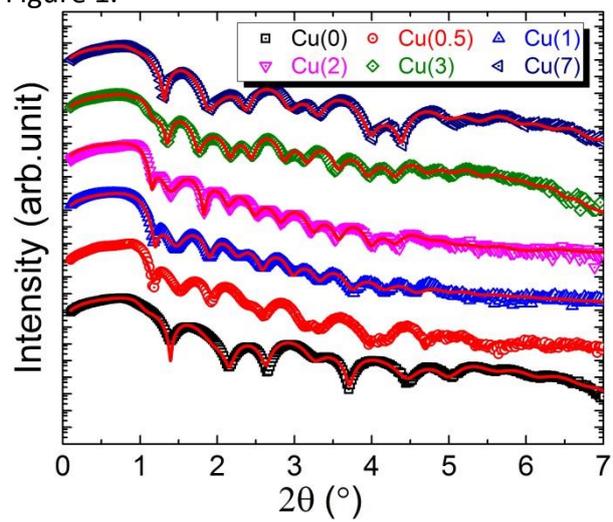

Figure 1. XRR spectra for the Cu(0), Cu(0.5), Cu(1), Cu(2), Cu(3) and Cu(7) samples. Symbols are experimental data and red solid lines are fits to the experimental data.

Figure 2.

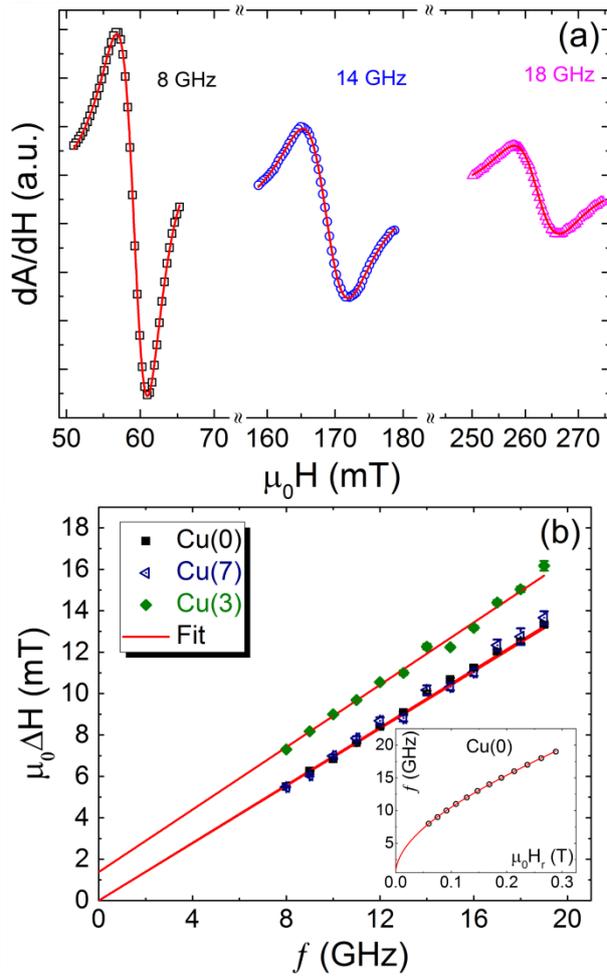

Figure 2. (a) *In-plane* FMR spectra of Cu(0) at different frequencies. (b) $\mu_0\Delta H$ vs. $f$ for the Cu(0), Cu(3) and Cu(7) samples. The inset shows $f$ vs. $\mu_0 H_r$ for the Cu(0) sample. The solid lines are fits to the experimental data as described in the text.

Figure 3.

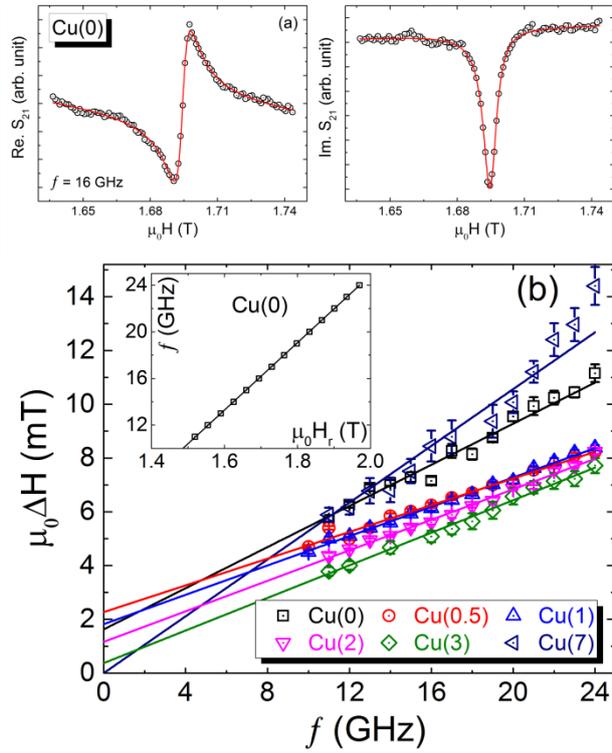

Figure 3. (a) Real and imaginary components of the *out-of-plane* FMR spectrum for the Cu(0) sample at 16 GHz. (b) $\mu_0 \Delta H$ vs. $f$ for the Cu(0), Cu(0.5), Cu(1), Cu(2), Cu(3) and Cu(7) samples. The inset shows $f$ vs. $\mu_0 H_r$ for the Cu(0) sample. The solid lines are fits to the experimental data as described in the text.

Figure 4.

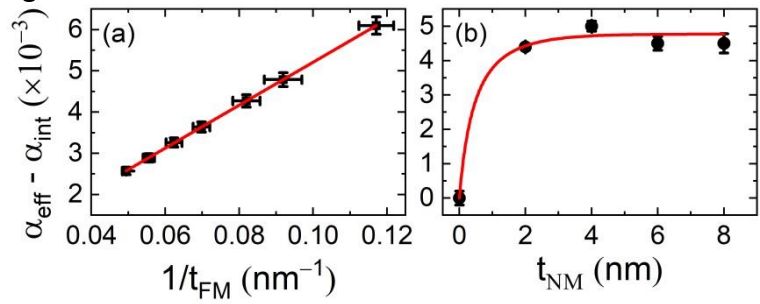

Figure 4. (a) $\alpha_{eff}$ vs. $1/t_{FM}$ and (b) $\alpha_{eff}$ vs. $t_{NM}$ for the CFA($t_{FM}$)/β-Ta($t_{NM}$) series thin films. The red solid lines are fits to the experimental data according to Eqs. (5-7).

Figure 5.

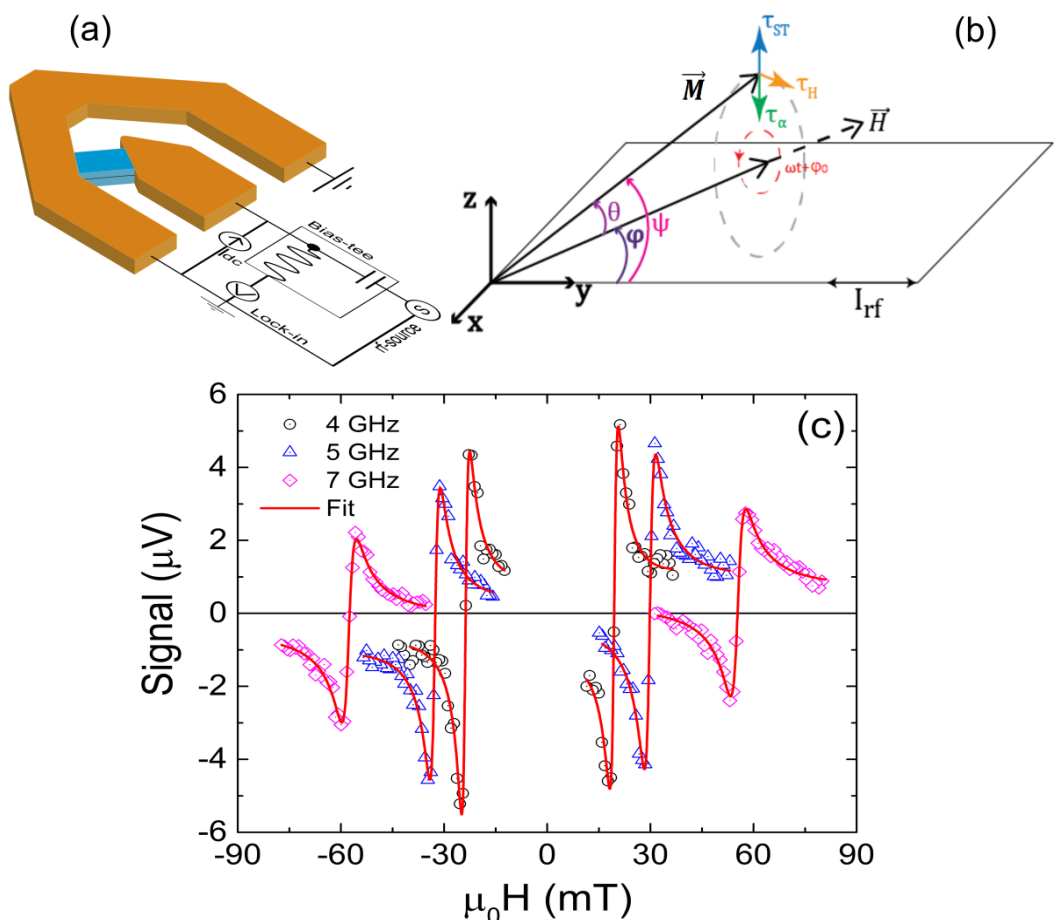

Figure 5. (a) Schematic of patterned structure and the ST-FMR measurement setup. (b) Schematic of spin torque induced magnetization $\vec{M}$ precession around its equilibrium direction at the driving frequency $\omega$ and phase delay, $\varphi_0$. $\psi$ is the angle of $\vec{M}$ with respect to the XY plane, and $\theta$ is the cone angle. $\tau_\alpha$, $\tau_{ST}$, and $\tau_H$ are damping-like torque, anti-damping spin-torque, and field-like torque, respectively; for details see Ref. [14]. (c) ST-FMR spectra for the Cu(0) sample in positive/negative magnetic field scans at different frequencies. The red solid lines are fits to the spectra using Eq. (8).

Figure 6.

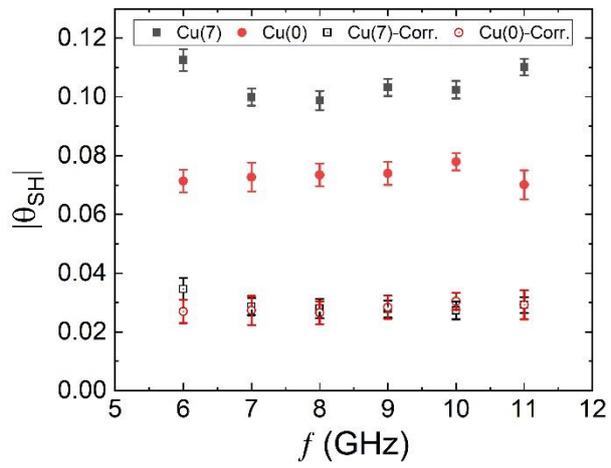

Figure 6. $|\theta_{SH}|$ vs. $f$ for the Cu(7) and Cu(0) samples with and without spin pumping correction.

Figure 7.

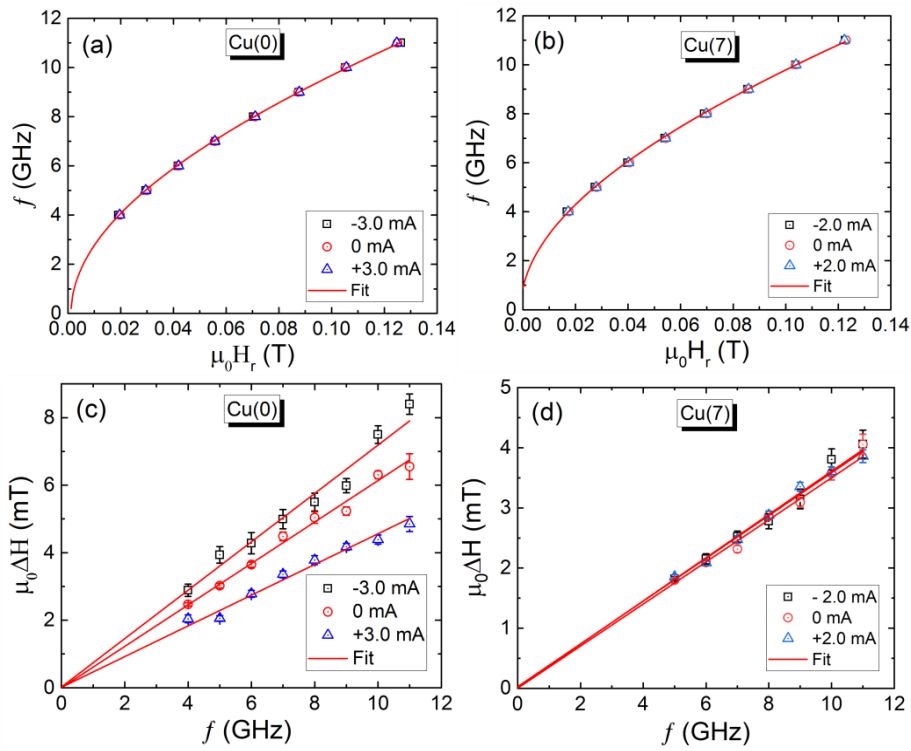

Figure 7. ST-FMR measured $f$ vs. $\mu_0 H_r$ for (a) Cu(0), (b) Cu(7), and $\mu_0 \Delta H$ vs. $f$ for (c) Cu(0), (d) Cu(7). The red solid lines are fits to the experimental data.

Figure 8.

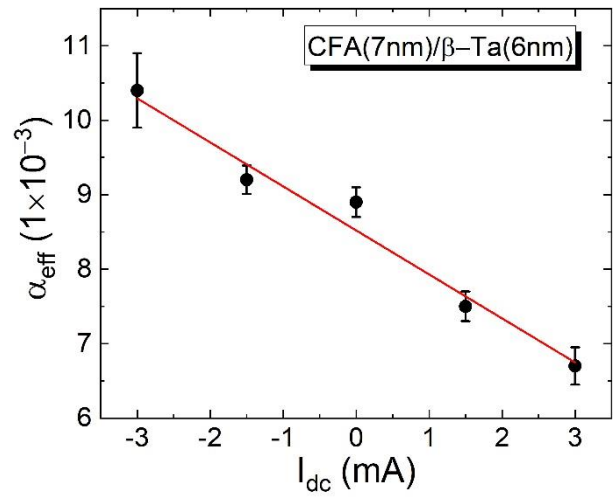

Figure 8. ST-FMR measured effective damping parameter ($\alpha_{eff}$) vs. $I_{dc}$ for the Cu(0) sample. The solid line is a linear fit to the experimental data.